\documentclass[aps,pre,preprint,superscriptaddress]{revtex4-1}
\usepackage{graphicx,dcolumn,bm}
\usepackage{amssymb,amstext,amsmath}
\usepackage{float}
\begin{document}


\title{Simple property of heterogeneous aspiration dynamics: Beyond weak selection}


\author{Lei Zhou}
\affiliation{Center for Complex Systems and Control, College of Engineering, Peking University, Beijing 100871, China}
\affiliation{Department of Ecology and Evolutionary Biology, Princeton University, Princeton, New Jersey 08544, USA}
\author{Bin Wu}
\affiliation{School of Sciences, Beijing University of Posts and Telecommunications, Beijing 100876, China}
\author{V\'{i}tor V.~Vasconcelos}
\affiliation{Department of Ecology and Evolutionary Biology, Princeton University, Princeton, New Jersey 08544, USA}
\author{Long Wang}
\email[]{longwang@pku.edu.cn}
\affiliation{Center for Complex Systems and Control, College of Engineering, Peking University, Beijing 100871, China}

\date{\today}

\begin{abstract}
    How individuals adapt their behavior in cultural evolution remains elusive.
    Theoretical studies have shown that the update rules chosen to model individual decision making can dramatically modify the evolutionary outcome of the population as a whole.
    This hints at the complexities of considering the personality of individuals in a population, where each one uses its own rule.
    Here, we investigate whether and how heterogeneity in the rules of behavior update alters the evolutionary outcome.
    We assume that individuals update behaviors by aspiration-based self-evaluation and they do so in their own ways.
    Under weak selection, we analytically reveal a simple property that holds for any two-strategy multi-player games in well-mixed populations and on regular graphs: the evolutionary outcome in a population with heterogeneous update rules is the weighted average of the outcomes in the corresponding homogeneous populations, and the associated weights are the frequencies of each update rule in the heterogeneous population.
    Beyond weak selection, we show that this property holds for public goods games.
    Our finding implies that heterogeneous aspiration dynamics is additive.
    This additivity greatly reduces the complexity induced by the underlying individual heterogeneity.
    Our work thus provides an efficient method to calculate evolutionary outcomes under heterogeneous update rules.
\end{abstract}


\maketitle

\section{Introduction}
How cooperative behavior evolves has puzzled researchers for decades.
The prisoner's dilemma \cite{Flood1950}, snowdrift game \cite{Sugden1986SG}, and stag-hunt game \cite{Schelling1980, Skyrms2004} received much attention in previous studies \cite{Nowak1992, Killingback1998, Hauert2004, Santos2006}.
They serve as the classical metaphors to study the evolution of cooperation in dyadic interactions.
Yet, in real-life situations, individuals are often involved in strategic interactions in larger groups, which can be captured by multi-player games \cite{Perc2013, Gokhale2010, Wu2013a}.
For example, environmental issues like deforestation, air pollution, and climate change are all public goods problems whose solutions usually need the collective action of more than two participants \cite{Milinski2008, Du2012, Vasconcelos2014}.
When studying cooperation between multiple players, a classical paradigm is the public goods game \cite{Groves1977, Ledyard1997}.
In this game, each individual can choose either to donate some amount to the common pool or to do nothing.
The total donation is then multiplied by an enhancement factor and the resulting benefit is equally distributed to all the individuals, irrespective of what they do.
Individuals therein have incentives to reap the benefit provided by others without donating.
This depicts the omnipresent social dilemma in which the interests of individuals and of the collective are in conflict.
Without proper regulation, this conflict often drives the population into the tragedy of commons \cite{Hardin1968} .

Evolutionary game theory provides a suitable framework to explore how cooperation emerges and persists under the above social dilemmas not only in genetic evolution \cite{MaynardSmith1973, MaynardSmith1982} but also in human cultural evolution \cite{Young1993, Borgers1997, Fudenberg1998}.
One key component of this framework is the underlying microscopic process (i.e., update rule), which determines how strategies spread over the population or how individuals adapt their behavior over time.
In the context of biological evolution, this is usually modeled by reproduction, inheritance, and replacement \cite{Nowak2004, Ohtsuki2006}.
Analogously, when studying strategic interactions in cultural evolution, imitation-based update rules are typically employed \cite{Szabo1998, Hauert2004, Traulsen2006b, Santos2006, Wu2015}.
These rules assume individuals copy the strategy of more successful peers by comparing payoffs, which relies on social information.
They create valuable criteria used by decision-makers in circumstances where rationality is bounded and shortcuts for decision-making are needed.
However, imitation-based rules are far from adequately depicting individuals' decision process, for instance, when social information is unreliable upfront.
Another heuristics, which is commonly found in both animal and human behavioral ecology, is the aspiration-driven decision-making.
Aspiration-driven decision-making assumes that individuals depend on personal rather than social information to make decisions: they self-evaluate behaviors by comparing payoffs with their endogenous aspirations and switch if the performance is not good enough.
For example, bumblebees most often stay and probe another flower of a plant when the previously probed one at the same plant has a larger volume of nectar than a threshold, otherwise they leave immediately \cite{Hodges1985};
honeybee, \textit{Apis mellifera}, and Norway rat, \textit{Rattus norvegicus}, are found to follow a copy-if-dissatisfied foraging strategy \cite{Galef2008, Gruter2013}.
As for humans, the ubiquity of reference points \cite{Bendor2011} and satisficing strategies \cite{Simon1947, Simon1959, Brown2004} clearly implies an underlying aspiration-based decision-making heuristic.

The prevalence of these two classes of rules elicits intensive studies on how they shape the evolution of cooperation \cite{Nowak1992, Nowak2004, Szabo2007, Chen2008, Du2014, Du2015}.
Although interesting phenomena are revealed, a tacit assumption in these studies is that all individuals use the same update rule.
Indeed, models with heterogeneous update rules are much harder to analyze than their homogeneous counterparts due to the high dimensionality and increasing complexity.
Nonetheless, recent behavioral experiments indicate that humans have consistent individual differences on how they gather information and process it to make decisions \cite{Worthy2013, Van2015}.
In addition, the heterogeneity of decision-making is suggested to be vital for understanding human strategic behavior \cite{Grujic2010}.
More importantly, incorporating individual variations into the models may result in different predictions when compared with those obtained under the assumption of homogeneity \cite{Vasconcelos2014, Molleman2014}.
To better understand the pattern resulting from individual strategic interactions, it is thus necessary to incorporate heterogeneity into update rules and try to deal with the rising complexity.
Heterogeneity in update rules can be modeled by assigning different individuals different types of update rules \cite{Moyano2009, Cardillo2010} or the same type of update rule but realized by different (update) functions \cite{Kirchkamp1999, Szabo2009, Szolnoki2009, Wu2013}.
For example, the pioneering work by Kirchkamp \cite{Kirchkamp1999} explores the evolution of update rules characterized by functions with three parameters.
The evolutionary dynamics with a mixture of individuals using different types of imitation-based rules is also investigated \cite{Moyano2009}.
However, most of these results are obtained by simulations and there is no general property connecting the heterogeneous population with its homogeneous counterparts.
A recent study which explores the mixing of innovative and imitative dynamics indeed analyzes the possible connections between heterogeneous and homogeneous populations but they fail to reveal any general property \cite{Amaral2018}.

Here, we propose heterogeneous aspiration dynamics, where each individual adopts aspiration-based rules with individualized update functions.
Our aim is to explore (i) whether and how the heterogeneity in update rules alters the evolutionary outcome, and (ii) the relation between the evolutionary outcome of the heterogeneous population and those of its homogeneous counterparts.
Resorting to tools stemming from statistical physics, we first derive the deterministic equations for heterogeneous aspiration dynamics in both well-mixed and structured populations.
We find that the solutions to these equations agree very well with evolutionary outcomes calculated from simulations.
Then for both weak and strong selections, we reveal a simple property of heterogeneous aspiration dynamics in public goods games: the evolutionary outcome of a heterogeneous population is the weighted average of the outcomes of the corresponding homogeneous populations, and the associated weights are the frequencies of each update rule in the heterogeneous population.
This implies that heterogeneous aspiration dynamics is additive.
By virtue of this property, one can greatly reduce the computational complexity involved in the original multi-dimensional birth-death process, which is in general analytically intractable.

This paper is organized as follows.
In Sec.~\ref{sectionModel}, we present our model about heterogeneous aspiration dynamics.
In Sec.~\ref{sectionResults}, we briefly explain our methods and derive the set of equations under any selection intensity in both well-mixed and structured populations.
In Sec.~\ref{AdditivityWeakWM}, we derive the condition for strategy (abundance) dominance and reveal a simple property (i.e., additivity) that holds for any two-strategy multiplayer games under weak selection.
We also perform simulations to validate our results on the condition for strategy dominance.
In Sec.~\ref{AdditivityStrong}, we compare numerical solutions with simulations to show that our equations accurately predict the evolutionary outcomes for a large range of selection intensities beyond weak selection.
Meanwhile, we find that additivity applies to public goods games under strong selection intensities.
In Sec.~\ref{sectionDiscussion}, our findings are summarized and we offer some discussion.
\section{Model}\label{sectionModel}
\begin{figure}[!ht]
  \centering
  \includegraphics[width=\textwidth]{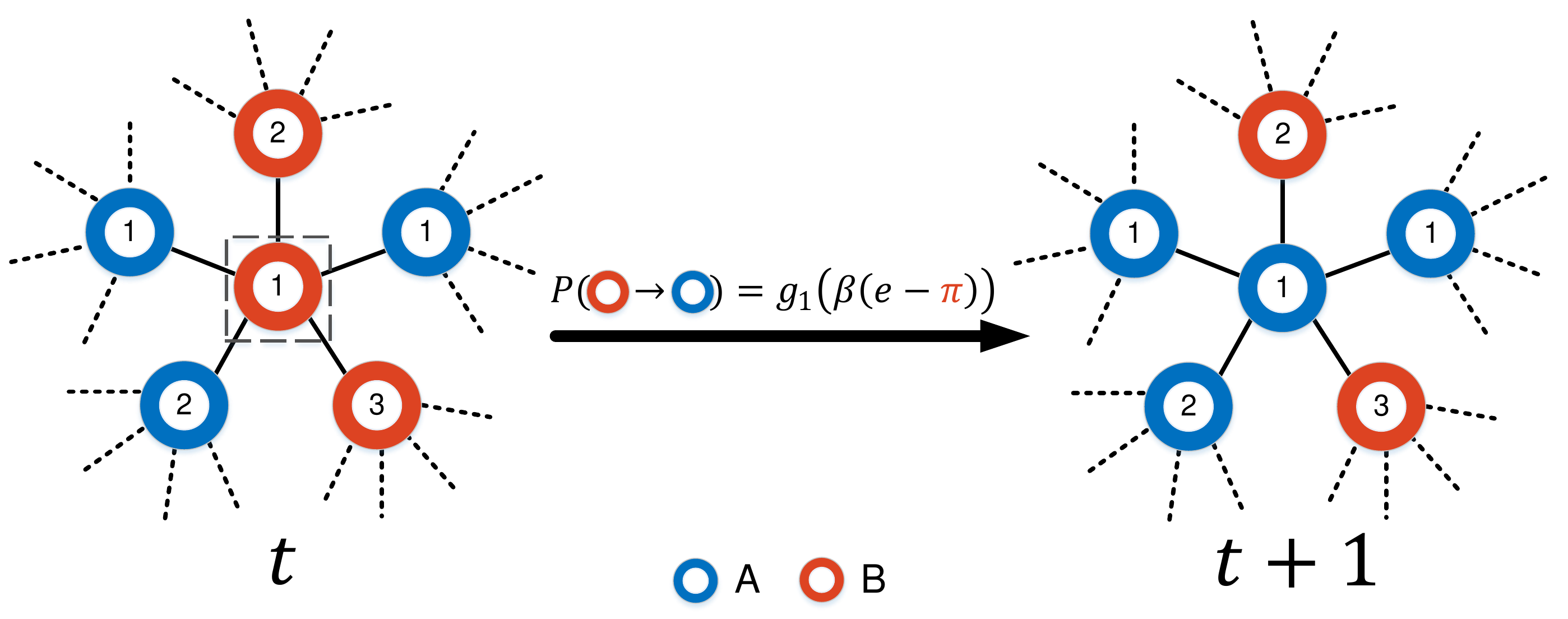}\\
  \caption{Schematic of the heterogeneous aspiration dynamics.
  The population structure is depicted by a network, where nodes are occupied by individuals and links represent social ties.
  Initially, each individual is assigned to a strategy (the color of the outer circle) and an update function (the number in the inner circle).
  At each time step, a focal individual is randomly selected to revise its strategy using its own update function.
  As illustrated, at time $t$, an individual playing strategy $B$ (red circle) is chosen to update (marked by the dashed square).
  It plays games with its neighbors and collects a payoff $\pi_B$.
  With the probability given by its own update function, $g_1(\beta(e-\pi_B))$, it switches to another strategy (here, strategy $A$, blue circle).
  Note that even if all neighbors of the focal individual use strategy $B$, it is still possible for the focal individual to switch to strategy $A$ since aspiration-based update rules are innovative \cite{Szabo2007, Amaral2018}.
  According to the definition of update functions, an individual's tendency to change its strategy will decrease as its associated payoff increases.
  The aspiration level $e$ serves as a benchmark for evaluating the performance of strategies and individuals' satisfaction \cite{Simon1959}.
  Moreover, the strictness of this evaluation is controlled by the selection intensity $\beta$.
  If $\beta \rightarrow \infty$, individuals switch deterministically if it is not satisfied.
  If $\beta \rightarrow 0$, their payoffs have a small impact on their decision-makings and they will switch strategies with a nearly constant probability.
    }
  \label{fig1}
\end{figure}

In our model, the population size is $N$.
Individuals play $d$-player games with others.
In the game, they can choose to play either strategy $A$ or strategy $B$.
We specify that an $A$ ($B$) player facing $k$ other $A$ coplayers among the rest $d-1$ coplayers will receive a payoff $a_k$ ($b_k$). The corresponding payoff matrix is given by the table below.
\begin{table}[H]
    \centering
    \caption{Payoff matrix of a $d$-player game.}
  \begin{tabular}{cccccccc}
    \hline
    \hline
    Number of $A$ opponents & 0 & 1 & 2 & $\cdots$ & $k$ & $\cdots$ & $d-1$ \\
    \hline
    $A$ & $a_0$ & $a_1$ & $a_2$ & $\cdots$ & $a_k$ & $\cdots$ & $a_{d-1}$\\
    $B$ & $b_0$ & $b_1$ & $b_2$ & $\cdots$ & $b_k$ & $\cdots$ & $b_{d-1}$\\
    \hline
    \hline
  \end{tabular}
  \label{payoffMatrixGeneral}
\end{table}
For the strategy updating, there are $M$ ($M \ll N$) update functions $g_i(u)$ ($i=1,2, \cdots,M$) in the population.
Here, we assume for any individual $X$, $u = \beta(e-\pi_X)$ where $\pi_X$ is its average payoff, $e$ the aspiration level, and $\beta$ the selection intensity (see Fig.~\ref{fig1} for further explanation).
These aspiration-based update functions $g_i\left(u\right)$ map the difference between aspiration and payoff into a probability, with which individuals switch to another strategy (for two-strategy games, the switching is either from strategy $B$ to $A$ or from $A$ to $B$).
Different update functions characterize individuals' personalities on the decision-making process.
For example, individuals using the update function $1/(1+\exp(-u))$ are more likely to switch than those using $1/(1+10\exp(-u))$.
For convenience, we will use $g_i$, $g_i(u)$ and $g_i\left(\beta(e-\pi_X)\right)$ interchangeably throughout this paper.

In our model, each individual is equipped with one of the $M$ update functions and the fraction of individuals using $g_i$ is denoted as $z_i$.
We assume that individuals do not change their update functions during the evolution.
Besides, each update function $g_i(u)$ should satisfy the following constraints:
\begin{enumerate}
  \item it is a probability, i.e., $g_i(u) \in [0,1]$ for any $u\in(-\infty,+\infty)$;
  \item it is a monotonically increasing function of $u$, i.e. $g_i'(u)>0$ for any $u$;
  \item $g_i(0)>0$.
\end{enumerate}
The first constraint is self-explanatory.
The second one ensures that the update rules are evolutionary, which means that when individuals get higher payoffs they should have a decreasing tendency to switch their strategies.
In this sense, the strategy that generates a higher (lower) payoff is more likely to be kept (discarded) in the population.
The third one prevents the frozen dynamics in the neutral case when $\beta=0$.

For the heterogeneous aspiration dynamics, at each generation, a focal individual $X$ is randomly selected from the population.
It collects its payoff $\pi_X$ by engaging in $d$-player games.
In a well-mixed population, the other $d-1$ players are randomly sampled from the rest of population \cite{Hauert2006, Zhou2018}.
In a structured population depicted by a $(d-1)$-regular graph, each individual organizes a game including itself and its $d-1$ nearest neighbours.
Thus, individual $X$ participates in a total of $d$ games organized by itself and its $d-1$ nearest neighbors \cite{Li2015, Zhou2015}.
After the games, the focal individual $X$ compares its payoff $\pi_X$ with its aspiration $e$.
Based on its own update function $g_{i_X}$ ($i_X$ is the index of $X$'s update function), it chooses to switch to the other strategy with probability $g_{i_X}$ or keep the same strategy with the complementary probability.
Noteworthy, this implies that heterogeneous aspiration dynamics allows individuals to switch to strategies absent in their neighborhood, which indicates that aspiration-based rules are innovative \cite{Szabo2007, Amaral2018}.

Generally, the above process can be modeled by a Markov chain with $2^N$ states where each state specifies which individual uses what strategy.
This Markov chain is aperiodic and irreducible, admitting a unique stationary distribution \cite{Wu2018}.
In this distribution, we calculate the average frequencies (i.e., abundances) of strategy $A$ and $B$.
If the average frequency of strategy $A$ is greater than that of $B$, we say that strategy $A$ outcompetes $B$ in abundance; otherwise, strategy $B$ outcompetes $A$ in abundance \cite{Tarnita2009, Wu2018}.
\section{Results}\label{sectionResults}

\begin{table}[htbp]
\caption{Main notations used in the Results Section. The upper part contains fixed parameters while the lower one the dynamic variables.}
\begin{tabular}{lcll}
\hline \hline
Notation&~~~~~~&Definition\\
\hline
$N$         &~~~~~~&\text{Population size}  \\
$d$         &~~~~~~&\text{Group size} \\
$k$         &~~~~~~&\text{Degree of regular graphs, }$k=d-1$ \\
$\beta$     &~~~~~~&\text{Selection intensity} \\
$e$         &~~~~~~&\text{Aspiration level}  \\
$g_i$       &~~~~~~&\text{Update function $i$} \\
$z_i$       &~~~~~~&\text{The fraction of individuals using $g_i$ in the population.  $0<z_i<1$} \\
$N_i$       &~~~~~~&\text{The number of individuals using $g_i$ in the population. $N_i=z_i N \gg 1$} \\
\hline
$n_i$       &~~~~~~&\text{The number of $A$-players in the well-mixed population}\\
$y_i$       &~~~~~~&\text{The frequency of $A$-players in the well-mixed population.  $y_i=n_i/N$} \\
$x_i$       &~~~~~~&\multicolumn{2}{p{14cm}}{\raggedright The frequency of $A$-players among the individuals who use $g_i$ in the well-mixed population. $x_i=n_i/N_i=y_i/z_i$}\\
$p_{A_i}$ &~~~~~~&\multicolumn{2}{p{14cm}}{\raggedright The frequency of $A$-players among the individuals who use $g_i$ on regular graphs}\\
$q_{A|A}$ &~~~~~~&\multicolumn{2}{p{14cm}}{\raggedright The conditional probability of finding an $A$-player in a focal individual's neighborhood, given that the focal individual is an $A$-player}\\
$\pi_A, \pi_B$ &~~~~~~&\text{Average payoffs of $A, B$-players} \\
\hline \hline
\end{tabular}
\label{notationTable}
\end{table}
We consider the simplest case where there are $M=2$ update functions present, $g_1$ and $g_2$.
In the population, the fractions of individuals using them are $z_1$ and $z_2$, respectively.
Here, both $z_1$ and $z_2$ are positive constants and they sum up to 1.
Accordingly, the number of individuals using $g_1$ and $g_2$ are $N_1 = z_1 N \gg 1$ and $N_2 = z_2 N \gg 1$.
As mentioned in the previous section, individuals do not change their update functions during the evolution, which means that $z_i$ and $N_i$ are \textit{fixed} parameters ($i=1,2$).
Meanwhile, since individuals keep revising their strategies, the frequency of individuals playing strategy $A$ (i.e., $A$-players) changes over time (i.e., $x_i$, $y_i$, and $p_{A_i}$ in Table \ref{notationTable}).

Under these settings, we derive deterministic equations in the large $N$ limit ($N\rightarrow \infty$) for both well-mixed populations (see Appendix \ref{AppendixWellMixed}) and structured ones represented by regular graphs (see Appendix \ref{AppendixStructured}).
In a nutshell, in well-mixed populations, following a similar procedure to that in \cite{Traulsen2006, Pacheco2014, Vasconcelos2017}, we first obtain the Fokker-Planck equation by a Kramers-Moyal expansion and then the corresponding stochastic differential equations.
After that, by taking the limit $N\rightarrow \infty$, we obtain the set of deterministic equations.
In structured populations, to capture the additional spatial correlation, we use the method of pair approximation \cite{Ohtsuki2006}.

Let us first consider well-mixed populations.
We denote the number of $A$-players using update function $i$ as $n_i$ ($0 \leq n_i \leq N_i$, $i=1,2$).
Then the associated frequencies of $A$-players using update function $i$ are $y_i=n_i/N \in [0,z_i]$.
After some calculations (see details in Appendix \ref{AppendixWellMixed}), we reach the following deterministic equations
\begin{equation}\label{WellMixedODE}
    \dot{y}_i = (z_i-y_i)g_i\left(\beta(e-\pi_B(y_1,y_2))\right) - y_i g_i\left(\beta(e-\pi_A(y_1,y_2))\right)
\end{equation}
where $\pi_A(y_1,y_2) = \sum_{k=0}^{d-1}{{d-1}\choose k}(y_1+y_2)^k\left(1 - y_1-y_2\right)^{d-1-k}a_k$ and $\pi_B(y_1,y_2) = \sum_{k=0}^{d-1}{{d-1}\choose k}(y_1+y_2)^k\left(1- y_1-y_2\right)^{d-1-k}b_k$.
Note that the homogeneous case, where only $g_2$ or $g_1$ is present, is recovered from Eqs.~(\ref{WellMixedODE}) by setting $z_1 = 0$ ($y_1 \equiv 0$) or $z_2 = 0$ ($y_2 \equiv 0$).

For positive $z_1$ and $z_2$, we can normalize the variables $y_i$ to the range $[0,1]$ by introducing the new variables $x_i = y_i / z_i$.
Here, $x_i$ is the frequency of $A$-players among the individuals who use update function $g_i$.
Therefore, the equations governing the evolution of $x_i$ in well-mixed populations are
\begin{equation}\label{NormalizedWellMixedODE}
    \dot{x}_i = (1-x_i)g_i\left(\beta(e-\pi_B(x_1,x_2))\right) - x_i g_i\left(\beta(e-\pi_A(x_1,x_2))\right)
\end{equation}
where
\begin{eqnarray*}
  \pi_A(x_1,x_2) &=& \sum_{k=0}^{d-1}{{d-1}\choose k}(z_1x_1+z_2x_2)^k\left(1 - z_1 x_1 -z_2 x_2\right)^{d-1-k}a_k,  \\
  \pi_B(x_1,x_2) &=& \sum_{k=0}^{d-1}{{d-1}\choose k}(z_1x_1+z_2x_2)^k\left(1-z_1 x_1 - z_2 x_2\right)^{d-1-k}b_k.
\end{eqnarray*}

Now we start to derive the equations for structured populations.
To do this, we tailor the pair approximation method for the heterogeneous aspiration dynamics.
Following the convention in \cite{Ohtsuki2006}, some notations are introduced here.
The degree of the regular network is $k = d-1$.
The frequency of strategy $A$ in the population is $p_A$ and that of strategy $B$ is $p_B$.
The probability to find a $YZ$ pair is denoted as $p_{YZ}$ and
the conditional probability for a $Y$ individual to find a $Z$ neighbor is denoted as $q_{Z|Y}$ ($Y, Z =A,B$).
Within the individuals using update function $g_i$, the frequency of strategy $A$ is $p_{A_i}$ and that of strategy $B$ is $p_{B_i}$.
The relationship between these notations are $p_A+p_B=1$, $p_{A_i}+p_{B_i}=1$, $p_Y=z_1p_{Y_1}+z_2p_{Y_2}$, $p_{YZ}= p_Y\cdot q_{Z|Y}$ ($i=1,2$ and $Y, Z =A,B$).

After the calculations, we obtain the equations for heterogeneous aspiration dynamics in structured populations (see detailed derivation in Appendix \ref{AppendixStructured}).
To simplify notations, we define $Q_A(k,k_A)={k \choose k_A}q_{A|A}^{k_A} q_{B|A}^{k-k_A}$ as the probability of finding $k_A$ $A$-players among the $k$ neighbors of a focal individual, given that this focal individual uses strategy $A$; similarly, $Q_B(k,k_A)={k \choose k_A}q_{A|B}^{k_A} q_{B|B}^{k-k_A}$ is the probability of finding $k_A$ $A$-players in the neighborhood of an individual using strategy $B$.
Then, the heterogeneous aspiration dynamics in structured populations is described by the following equations
\begin{align}
\dot{p}_{A_1} =& \frac{p_{B_1}}{N}\sum_{k_A=0}^k Q_B(k,k_A)~g_1\left(u_{B, k_A}\right) - \frac{p_{A_1}}{N}\sum_{k_A=0}^k Q_A(k,k_A)~g_1\left(u_{A, k_A}\right), \label{DpA1NewMain}\\
\dot{p}_{A_2} =& \frac{p_{B_2}}{N}\sum_{k_A=0}^k Q_B(k,k_A)~g_2\left(u_{B, k_A}\right)- \frac{p_{A_2}}{N}\sum_{k_A=0}^k Q_A(k,k_A)~g_2\left(u_{A, k_A}\right), \label{DpA2NewMain} \\
\dot{q}_{A|A} =& \frac{1}{p_A}\sum_{k_A=0}^k Q_B(k,k_A)\left(\frac{2k_A}{Nk}\right)\left[z_1p_{B_1}~g_1\left(u_{B, k_A}\right)+z_2p_{B_2}~g_2\left(u_{B, k_A}\right) \right] \nonumber\\
&- \frac{1}{p_A}\sum_{k_A=0}^k Q_A(k,k_A)\left(\frac{2k_A}{Nk}\right)\left[z_1p_{A_1}g_1\left(u_{A, k_A}\right)+z_2p_{A_2}g_2\left(u_{A, k_A}\right) \right] \nonumber\\
&  - \frac{q_{A|A}}{p_A}(z_1\dot{p}_{A_1}+z_2\dot{p}_{A_2}) \label{DqAANewMain},
\end{align}
where $u_{A,k_A} = \beta(e-\pi_{A,k_A})$, $u_{B,k_A} = \beta(e-\pi_{B,k_A})$, and
\begin{eqnarray*}
  \pi_{A,k_A} &=& \frac{1}{k+1}\left[a_{k_A}+k_{A}\sum_{l_A=0}^{k-1}Q_A(k-1,l_A)a_{l_A+1}+(k-k_A)\sum_{l_A=0}^{k-1}
  Q_B(k-1,l_A)a_{l_A}\right], \\
  \pi_{B,k_A} &=& \frac{1}{k+1}\left[b_{k_A}+k_{A}\sum_{l_A=0}^{k-1}Q_A(k-1,l_A)b_{l_A+1} +(k-k_A)\sum_{l_A=0}^{k-1}Q_B(k-1,l_A)b_{l_A}\right],
\end{eqnarray*}
(see \cite{Li2016}).
For a better understanding, we explain the terms in $\pi_{A,k_A}$ one by one: $a_{k_A}$ is the payoff derived from the game organized by the focal $A$-player itself; $k_A \sum_{l_A=0}^{k-1}Q_A(k-1,l_A)a_{l_A+1}$ is the total payoff gained by participating in the games organized by the $k_A$ $A$-neighbors; $(k-k_A)\sum_{l_A=0}^{k-1} Q_B(k-1,l_A)a_{l_A}$ is the total payoff obtained by engaging in the games organized by the rest $k-k_A$ $B$-neighbors.
Since $q_{B|A}=1-q_{A|A}$, $q_{A|B}=(z_1p_{A_1}+z_2p_{A_2})(1-q_{A|A})/(1 - z_1p_{A_1}-z_2p_{A_2})$, and $q_{B|B} = 1-q_{A|B}$, Eqs. (\ref{DpA1NewMain}-\ref{DqAANewMain}) depict a closed dynamic system with three state variables $p_{A_1}$, $p_{A_2}$, and $q_{A|A}$.

As shown above, for $M=2$ update functions, we need two variables ($x_1$ and $x_2$) to describe the dynamics of the system in well-mixed populations.
They are the frequencies of $A$-players among the individuals using update function $g_1$ or $g_2$.
On regular graphs, the same two variables are $p_{A_1}$ and $p_{A_2}$.
Meanwhile, to capture the spatial correlation resulting from the population structure, we need an additional variable $q_{A|A}$, which depicts the assortment of strategy $A$.
In general, for $M>2$ update functions with $N_i\gg 1$ for all $i$ ($i=1,2,\cdots,M$), $M$ independent variables are needed in well-mixed populations and $M+1$ ones on regular graphs.
To obtain the corresponding set of deterministic equations, it is straightforward to generalize our approach in Appendix \ref{AppendixWellMixed} and Appendix \ref{AppendixStructured}.

With the dynamical equations in well-mixed and structured populations, we now turn to analyze their long-term behavior.
We focus on the average abundance of strategy $A$ in the whole population in the steady state (i.e., in equilibrium).
Let us denote this quantity of interest as $x^*$ and the fixed points of Eqs.~(\ref{NormalizedWellMixedODE}) as $(x_1^*,x_2^*)$.
By definition, we have $x^* = z_1 x_1^* + z_2 x_2^*$ in well-mixed populations.
Similarly, in structured populations, $x^* = z_1 p_{A_1}^* + z_2 p_{A_2}^*$, where $p_{A_1}^*$ and $p_{A_2}^*$ are the first two coordinates of the fixed points for Eqs.~(\ref{DpA1NewMain}-\ref{DqAANewMain}).

In what follows, we will focus on the effect selection intensity, $\beta$, has on the long-term dynamics. In the limit of weak selection (i.e., $\beta \rightarrow 0$), the payoff and aspiration level have a small impact on individuals' decisions and they will switch to the other strategy with a probability close to $g(0)$; in the strong selection limit, $\beta \rightarrow \infty$, the difference between aspiration and payoff plays a decisive role in individuals' strategy updating: they will deterministically stick to their strategy when their aspiration is met and switch otherwise.
For an intermediate $\beta$, strategies generating high payoffs will be more likely to be repeated.

\subsection{Weak selection}\label{AdditivityWeakWM}
Under the weak selection limit $\beta \rightarrow 0$, in both well-mixed and structured populations, the average abundance of strategy $A$ in the steady state is
\begin{equation}\label{AvgAbundanceMixed}
  x^* =  \frac{1}{2}+ \frac{1}{2^{d+1}}\left(z_1\frac{g'_1(0)}{g_1(0)}+z_2\frac{g'_2(0)}{g_2(0)}\right)\sum_{k=0}^{d-1}{{d-1}\choose k}(a_k-b_k)\beta + O(\beta^2)
\end{equation}
and that of $B$ is $1-x^*$ (see Appendix \ref{AppendixWellMixed} and Appendix \ref{AppendixStructured} for detailed calculations).
Rearranging the items in the above equation, we have
\begin{eqnarray}
 x^* &\approx& z_1\underbrace{\left(\frac{1}{2}+ \frac{1}{2^{d+1}}\frac{g'_1(0)}{g_1(0)}\sum_{k=0}^{d-1}{{d-1}\choose k}(a_k-b_k)\beta+O(\beta^2)\right)}_{x_{\text{I}}^*} \nonumber\\
 & &+ z_2\underbrace{\left(\frac{1}{2}+ \frac{1}{2^{d+1}}\frac{g'_2(0)}{g_2(0)}\sum_{k=0}^{d-1}{{d-1}\choose k}(a_k-b_k)\beta+O(\beta^2)\right)}_{x_{\text{II}}^*} \label{weightedSum}.
\end{eqnarray}
As shown in \cite{Du2014}, $x_{\text{I}}^*$ ($x_{\text{II}}^*$) is exactly the average abundance of strategy $A$ in the steady state of the homogeneous population where all the individuals employ the same update function $g_1$ ($g_2$).
Since $z_1,z_2> 0$ and $z_1+z_2=1$, Eq.~(\ref{weightedSum}) indicates that under the limit of weak selection, $x^*\approx z_1 x_{\text{I}}^* + z_2 x_{\text{II}}^*$.
This means that the average abundance of strategy $A$ in the heterogeneous population is just the weighted average of those in the homogeneous ones.
In other words, aspiration dynamics with heterogenous update functions is additive, provided the selection intensity $\beta$ is sufficiently weak.

Moreover, since $g'_i(0)>0$ and $g_i(0)>0$ ($i=1,2$) in Eq.~(\ref{AvgAbundanceMixed}), we can derive the condition for strategy $A$ to outcompete strategy $B$ in abundance in the steady state (i.e., $x^* > 1/2$ \cite{Antal2009, Tarnita2009}).
This condition for strategy dominance is
\begin{equation}\label{conditionForStrategyA}
  \sum_{k=0}^{d-1}{{d-1}\choose k}(a_k-b_k) > 0.
\end{equation}
Note that the above condition is also found in a finite well-mixed population with a homogenous update function \cite{Du2014}.
Although condition (\ref{conditionForStrategyA}) is derived in infinite populations, our results will approximately hold for finite (but large) populations.
This is justified for two reasons: the binomial approximation remains good for large $N$; the stochasticity introduced by finite population size does not affect the average abundance of strategy $A$ since the fluctuation is Gaussian with a zero mean in the neighborhood of the steady state \cite{Van1992}.
Interestingly, condition (\ref{conditionForStrategyA}) indicates that under the weak selection limit, the heterogeneity of aspiration-based update functions does not affect the criterion to tell whether strategy $A$ is more abundant than strategy $B$.
Our finding thus extends the applicability of condition (\ref{conditionForStrategyA}) revealed in \cite{Du2014, Du2015} to the following scenarios: (i) mixed update functions in well-mixed populations, (ii) multi-player games on regular graphs, and (iii) mixed update functions and multi-player games on regular graphs.
This contrasts with the results for imitation-driven dynamics where the condition for strategy dominance derived under a mixture of birth-death and death-birth rules is not the same as its homogeneous counterparts, and it is sensitive to the frequency of each update rule in the population \cite{Zukewich2013}.

\begin{figure}[H]
  \centering
  \includegraphics[width=0.8\textwidth]{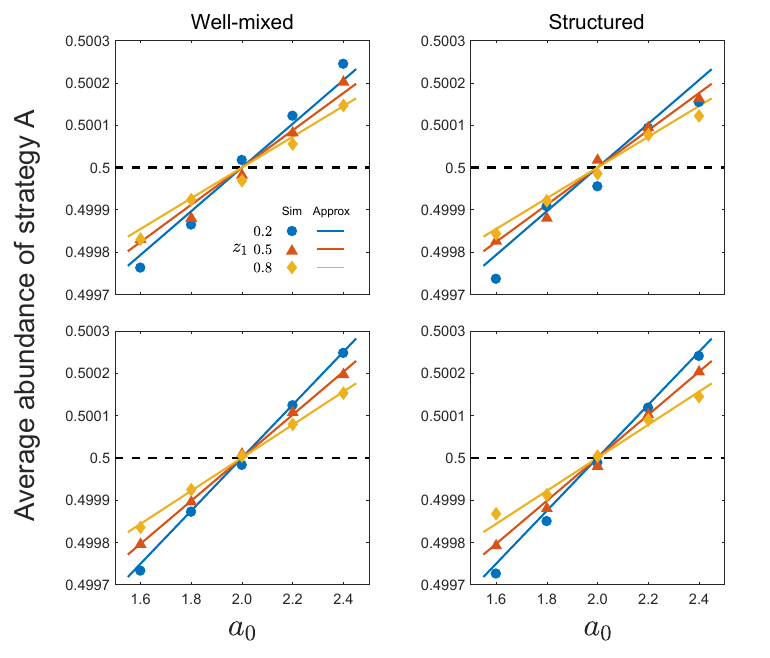}\\
  \caption{The condition for strategy $A$ to outcompete $B$ in abundance in three-player games for three fractions of update function $g_1$: $z_1=0.2$ (blue), $z_1=0.5$ (red), and $z_1=0.8$ (yellow).
  We plot the average abundance of strategy $A$ as a function of the payoff entry $a_0$.
  The other payoff entries are $a_1=2$, $a_2=1$, $b_0=4$, $b_1=1$, and $b_2=1$.
  The left panels are results in a well-mixed population and the right panels are the results in a structure population set up on a ring (i.e., 2-regular graph).
  For the average abundance of strategy $A$, analytical results (solid lines) predicted by Eq.~(\ref{AvgAbundanceMixed}) match the simulations (symbols).
  We test two kinds of mixture of update functions: the upper row for Fermi function $g_1(u)=1/(1+\exp(-u))$ and its variant $g_2(u)=1/(1+10\exp(-u))$; the lower row for $g_1(u)=1/(1+\exp(-u))$ and the rescaled error function $g_2(u)=(1+\text{erf}(u))/2$ where ${\rm erf}(u)=(2/\sqrt{\pi})\int_0^{u}\exp(-t^2)dt$ is the error function \cite{Wu2013}.
  The simulation results are obtained by averaging the mean abundance of strategy $A$ in 100 runs; in each run, this mean value is the average of the abundance of strategy $A$ in the last $3\times 10^7$ time steps after a transient time of $1\times 10^7$ time steps.
  Initially, we set the frequency of strategy $A$ to be 0.6.
  The selection intensity $\beta=0.01$.
  Other parameters: $N=500$, $d=3$, and $e=2.5$.}
  \label{fig2}
\end{figure}

To verify our analytical results, we implement agent-based simulations.
Here, we test the condition in three-players games with payoff entries $a_1=2$, $a_2=1$, $b_0=4$, $b_1=1$, and $b_2=1$, leaving $a_0$ as a tunable parameter.
For these games, applying condition (\ref{conditionForStrategyA}) immediately leads to the conclusion: if $a_0>2$, the average abundance of strategy $A$, $x^*$, is greater than one-half.
In Fig.~\ref{fig2}, we plot the average abundance of strategy $A$ obtained from simulations as a function of the payoff entry $a_0$.
Our results demonstrate that in both well-mixed and structured populations, for $a_0>2$, $x^* > 1/2$.
Moreover, this is true for different combinations of update functions, as shown in the upper and lower rows in Fig.~\ref{fig2}.
Furthermore, the average abundance of strategy $A$ predicted by Eq.~(\ref{AvgAbundanceMixed}) matches perfectly with the simulation results.
Our analytical results thus provide a simple and fast method to calculate the final evolutionary outcomes, which greatly reduces the computational cost involved in simulations.

\subsection{Strong selection}\label{AdditivityStrong}
In the previous section, we offer closed-form results on the final evolutionary outcomes as $\beta \rightarrow 0$.
These analytical results are shown to agree with simulation results very well for sufficiently small $\beta$ ($\beta=0.01$, in Fig.~\ref{fig2}).
Now we move further and extend our analysis to strong selection scenarios.
Under strong selection intensities, there are in general no closed-form solutions.
The reason is that we can no longer perform perturbation analysis at the neutral drift ($\beta=0$).
Only recently the evolutionary dynamics under strong selections was addressed analytically for special cases such as well-mixed populations \cite{Traulsen2006b, Wu2013} and rings \cite{Ohtsuki2006b, Van2012, Altrock2017}.
Ideally, the set of equations we derived for well-mixed (see Eqs.~(\ref{NormalizedWellMixedODE})) and structured (see Eqs.~(\ref{DpA1NewMain}-\ref{DqAANewMain})) populations apply to any selection intensity.
However, as selection intensity gets strong, the payoffs will greatly affect the switching behavior of individuals.
It is expected that the dynamical equations obtained from the pair approximation may generate predictions that largely deviate from the real ones \cite{Szabo2005, Szabo2007}.
The reason partly lies in the inaccurate estimation of payoffs by considering only pair correlations, especially for multi-player games.

\begin{figure}[H]
  \centering
  \includegraphics[width=0.8\textwidth]{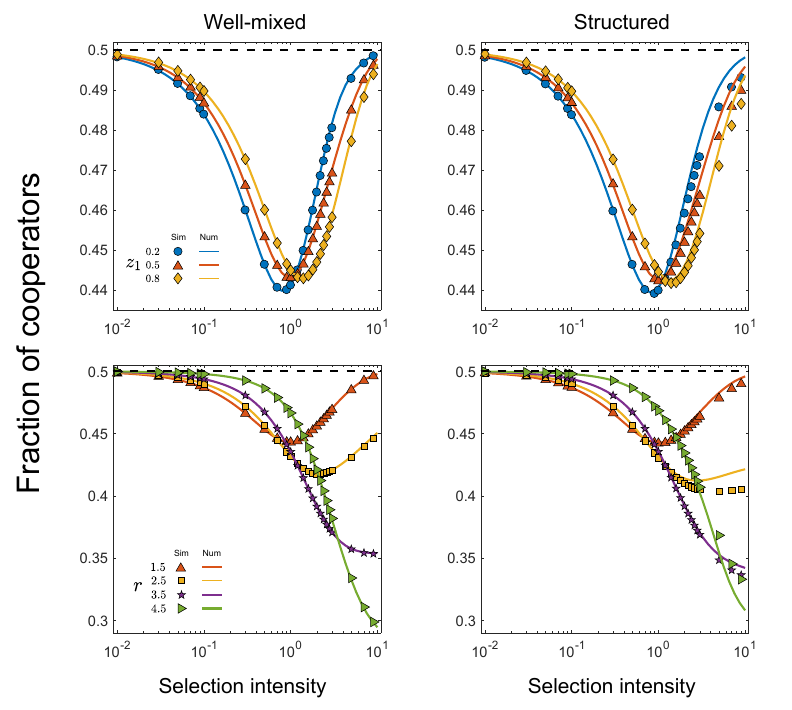}\\
  \caption{Mean fraction of cooperators in the five-player public goods game.
  The left panels are the results in a well-mixed population while the right panels are the results in a structured population set up on a square lattice.
  The symbols represent the simulation results and the solid lines the fixed points found by solving Eqs.~(\ref{NormalizedWellMixedODE}) for the well-mixed population and Eqs.~(\ref{DpA1NewMain}-\ref{DqAANewMain}) for the structured
  one.
  The payoff entries of the public goods game are $a_k=rc(k+1)/d-c$ and $b_k=rck/d$ ($k=0,1,\cdots, d-1$) where $r$ is the enhancement factor ($1<r<d$) and $c$ the cost of cooperation.
  In the upper row, different colors represents the results with $r=1.5$ under different $z_1$: $z_1=0.2$ (blue), $z_1=0.5$ (red), and $z_1=0.8$ (yellow).
  In the lower row, the results are obtained under $z_1=0.5$ with different $r$:
  $r=1.5$ (red), $r=2.5$ (yellow), $r=3.5$ (purple), and $r=4.5$ (green).
  We show that the numerical solutions are in agreement with the simulations for almost all the selection intensities considered here.
  The simulation results are obtained by averaging the mean fraction of cooperators in 50 runs; in each run, this mean value is the average of the fraction of cooperators in the last $3\times 10^7$ time steps after a relaxation time of $1\times 10^7$ time steps.
  Update function $g_1(u)=1/\left(1+\exp(-u)\right)$ and $g_2(u)=\left(1+{\rm erf}(u)\right)/2$ \cite{Wu2013}.
  The initial fraction of cooperators is set to be 0.4.
  Other parameters: $N=576$, $d=5$, $e=1.0$, and $c=1.0$.}
  \label{fig3}
\end{figure}

To test whether the equations we derive apply to strong selection scenarios and how good the predictions are, we employ the public goods games and show the average fraction of strategies for a large range of selection intensities, from weak ($\beta=0.01$) to strong ($\beta=9$).
In the context of public goods games, strategy $A$ is interpreted as \textit{cooperation} and $B$ \textit{defection}.
The payoff entries in Table \ref{payoffMatrixGeneral} become $a_k=rc(k+1)/d-c$ and $b_k=rck/d$ ($k=0,1,\cdots,d-1$), where $r$ is the enhancement factor ($1<r<d$) and $c$ the cost of cooperation.
In detail, we implement agent-based simulations and numerically calculate the fixed points of the set of Eqs.~(\ref{NormalizedWellMixedODE}) and (\ref{DpA1NewMain}-\ref{DqAANewMain}).
The stationary fraction of cooperators as a function of the selection intensity $\beta$ under different $z_1$ is plotted in the upper row of Fig.~\ref{fig3}.
Moreover, we fix $z_1=0.5$ and tune the enhancement factor $r$, which depicts the severity of the social dilemma, to see how our results change accordingly in the lower row of Fig.~\ref{fig3}.
The results in Fig.~\ref{fig3} show that the numerical solutions (solid lines) accurately predict the stationary fraction of cooperators for all the selection intensities considered here in well-mixed populations and for $\beta<3$ in structured ones.
In structured populations, when the selection intensity $\beta>3$, numerical solutions start to deviate from simulations.
As mentioned above, this is because strategy revisions are now largely affected by payoffs.
For public goods games on regular graphs, individuals' payoffs are affected not only by their nearest neighbors but also by their second-nearest neighbors.
The effect of the latter (i.e, triplet correlations) is neglected in the pair approximation, which leads to deviations.
Albeit this, the numerical solutions in structured populations match the simulations reasonably well for strong selections: the values are close to the ones obtained from simulations up to selection intensity $\beta=9$.
This suggests that for heterogenous aspiration dynamics, pair approximation can be an efficient method to calculate the evolutionary outcomes under strong selection intensities.

\begin{figure}[H]
  \centering
  \includegraphics[width=0.8\textwidth]{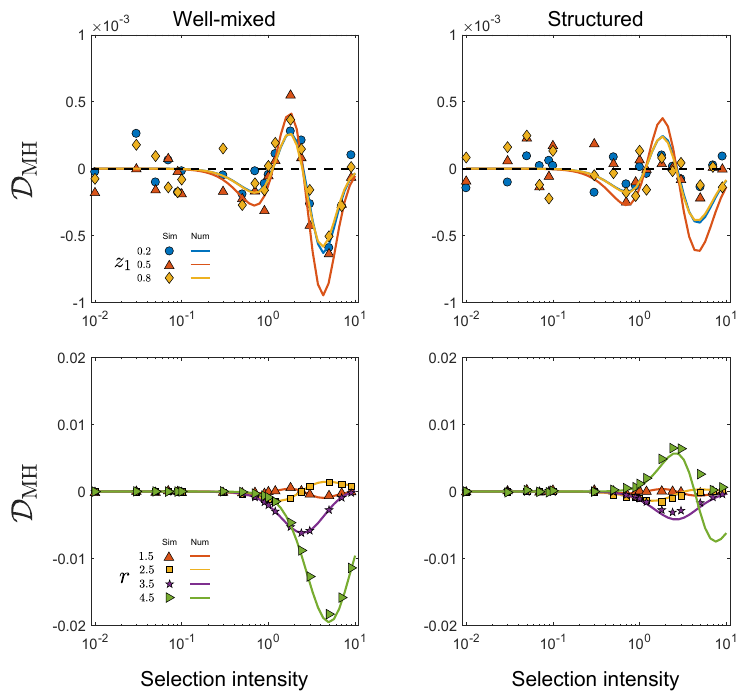}\\
  \caption{The additivity of aspiration dynamics with mixed (heterogeneous) update functions in public goods games.
  The stationary fraction of cooperators in the mixed population (with $g_1$ and $g_2$) is denoted as $x^*_{\text{mixed}}$ and that in the homogeneous ones is represented by $x^*_{\text{I}}$ (only with $g_1$) or $x^*_{{\text{II}}}$ (only with $g_2$).
  How well the results of the mixed system can be approximated by the weighted average of those in the homogeneous systems is measured by the difference $\mathcal{D}_{\text{MH}}=x^*_{\text{mixed}}-z_1 x^*_{\text{I}}-z_2 x^*_{\text{II}}$, where $z_i$ is the fraction of individuals using $g_i$ in the mixed population.
  We plot this difference as a function of the selection intensity $\beta$.
  Symbols represent results obtained from simulations while solid lines from numerical solutions.
  The upper row shows the results by fixing $r=1.5$ and varying $z_1$ while the lower row by fixing $z=0.5$ and varying $r$.
  The other notations and parameters are the same as those in Fig.~\ref{fig3}.
 }
  \label{fig4}
\end{figure}

In Sec.~\ref{AdditivityWeakWM}, under the limit of weak selection, we analytically derive the additivity of heterogeneous aspiration dynamics (see Eq.~(\ref{weightedSum})).
To assess how robust this property is to selection intensities, we calculate the difference between the average abundance of strategy $A$ in the population with mixed update functions and the weighted average of that in the homogeneous populations for various $\beta$.
Here, we denote this difference as $\mathcal{D}_{\text{MH}} = x^*_{\text{mixed}}-z_1 x^*_{\text{I}}-z_2 x^*_{\text{II}}$.
If $\mathcal{D}_{\text{MH}}=0$, the evolutionary outcome of the aspiration dynamics with mixed update functions can be perfectly estimated by the weighted average of those in the homogeneous populations.
If $\mathcal{D}_{\text{MH}}\neq0$, there are deviations in this estimation and the error is $|\mathcal{D}_{\text{MH}}|$.
Necessarily, if $\mathcal{D}_{\text{MH}}<0$, the weighted average $z_1 x^*_{\text{I}}+z_2 x^*_{\text{II}}$ overshoots the actual value $x^*_{\text{mixed}}$ of the mixed system; similarly, $\mathcal{D}_{\text{MH}}>0$ indicates  $x^*_{\text{mixed}}$ is instead underestimated.
In Fig.~\ref{fig4}, we plot $\mathcal{D}_{\text{MH}}$ calculated from simulations (symbols) and numerical solutions (solid lines) as a function of the selection intensity $\beta$.
The results shown in the upper row of Fig.~\ref{fig4} reveal that in both well-mixed and structured populations, $|\mathcal{D}_{\text{MH}}|$ reaches its peak when $z_1=0.5$ and $\beta$ is between 4 and 5.
It means that the discrepancy between $x_{\text{mixed}}^*$ and $z_1 x^*_{\text{I}}+z_2 x^*_{\text{II}}$ increases when $z_1$ and $z_2$ become closer to each other.
This may be caused by the increasing heterogeneity of the population when $z_1$ gets closer to $z_2$.
In the lower row, to investigate how the enhancement factor $r$ of the public goods game affects the evolutionary outcomes, we set $z_1=0.5$ and tune the value of $r$.
The results indicate that the peak of $|\mathcal{D}_{\text{MH}}|$ increases with $r$.
In the public goods game, for the parameters tested, we get that the maximum error of the approximation by the weighted average $z_1 x^*_{\text{I}}+z_2 x^*_{\text{II}}$ is less than $2\times 10^{-2}$.
This suggests that the additivity of aspiration dynamics with heterogeneous update functions still applies to strong selection.
Furthermore, the additivity seems to hold very well when $\beta<1$, where the maximum error is less than $2\times 10^{-3}$.
To find analytical explanations for our numerical analysis, we construct higher-order approximations (around $\beta=0$) for $\mathcal{D}_{\text{MH}}$ in well-mixed populations (see Appendix \ref{HigherOrderApprox}).
We obtain that for the public goods game (see Eq.~(\ref{DeviationHigherOrderApprox})),
\begin{equation*}
  \mathcal{D}_{\text{MH}} = \frac{1}{16}z_1z_2(d-1)\frac{rc}{d}\left(\frac{rc}{d}-c\right)^2G_1G_2\beta^3 + O(\beta^4),
\end{equation*}
where $G_1 = \frac{g_1'(0)}{g_1(0)}-\frac{g_2'(0)}{g_2(0)}$ and $G_2 = \left(\frac{g_2'(0)}{g_2(0)}\right)^2-\left(\frac{g_1'(0)}{g_1(0)}\right)^2+ \frac{g_1''(0)}{g_1(0)} - \frac{g_2''(0)}{g_2(0)}$.
This indicates $|\mathcal{D}_{\text{MH}}|$ starts to deviate from zero (i.e., perfectly additive) at $\beta^3$.
It explains the good preservation of additivity in public goods games for non-vanishing selection intensities.
The above equation also shows that the closeness between update function $g_1$ and $g_2$ affects $|\mathcal{D}_{\text{MH}}|$.

\begin{figure}[H]
  \centering
  \includegraphics[width=0.8\textwidth]{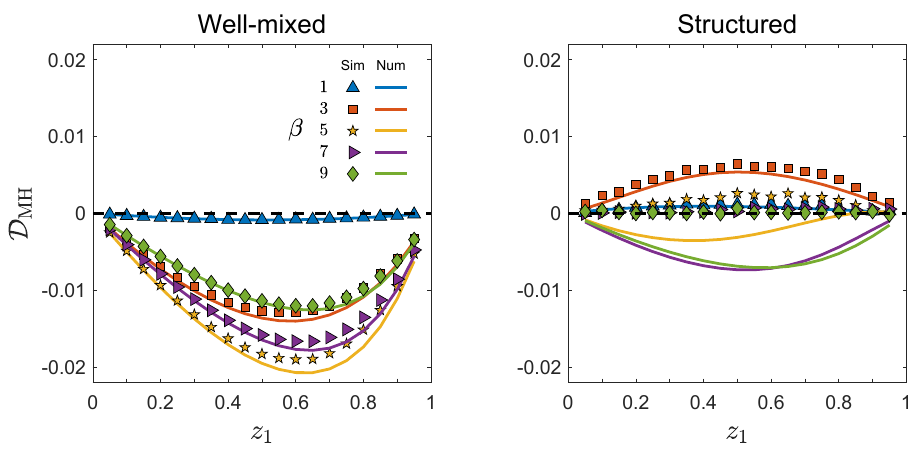}\\
  \caption{The robustness of additivity to the fraction $z_1$ of individuals using update function $g_1$.
  We plot $\mathcal{D}_{\text{MH}}$ as a function of $z_1$ for various (strong) selection intensities: $\beta=1$ (blue), $\beta=3$ (red), $\beta=5$ (yellow), $\beta=7$ (purple), and $\beta=9$ (green).
  The enhancement factor $r$ of the public goods game is set to be 4.5.
  Symbols represent results obtained from simulations while solid lines from numerical solutions.
  The other notations and parameters are the same as those in Fig.~\ref{fig3}.
 }
  \label{fig5}
\end{figure}

Up to now, we only test the additivity property of heterogeneous aspiration dynamics for a few values of $z_1$ ($z_1=0.2$, $0.5$, and $0.8$ in the upper row of Fig.~\ref{fig4}).
To test this property for a wider range of $z_1$, we plot $\mathcal{D}_{\text{MH}}$ as a function of $z_1$ under various strong selection intensities in Fig.~\ref{fig5}.
Our results show that both in well-mixed populations and on regular graphs, the additivity property is robust to $z_1$ (from $z_1=0.05$ to $z_1=0.95$).

Our numerical results are obtained based on the assumption of infinite population size.
However, the good agreement between numerical solutions and simulation results suggests that the additivity may still apply when the population size is finite, since all of our simulations are conducted in finite populations.
If this is proven to be true, the analytical tractability of the homogeneous aspiration  dynamics \cite{Du2014} can be readily utilized to approximate the results under two or more update functions.
Note that the multi-dimensional system induced by heterogeneous update functions is much more difficult to analyze and the stationary solution, in general, cannot be obtained analytically \cite{Van1992, Traulsen2006}.
In this sense, the additive property of heterogeneous aspiration dynamics as shown in Fig.~\ref{fig4} may greatly reduce the complexity induced by increasing dimensions, which saves a lot of computation time.

\section{Discussion and conclusions}\label{sectionDiscussion}
In this paper, we investigate how heterogeneous aspiration-based update functions affect the evolutionary outcome.
We show that aspiration dynamics with heterogeneous (mixed) update functions is additive.
This additivity means that the final evolutionary outcomes in a population with mixed update functions is the weighted average of the outcomes in the corresponding homogeneous populations.
Moreover, the associated weights are the frequencies of each update function in the mixed case.
Under the limit of weak selection, we analytically derive this additivity for any two-strategy multi-player games.
When the selection gets stronger, simulations and numerical results suggest this property still holds in public goods games.
Utilizing this property, we may circumvent the difficulty encountered in the analysis of multi-dimensional birth-death process and instead focus on the much simpler one-dimensional case.
Note that for the one-dimensional case with reflecting boundaries, the detailed balance is fulfilled and this makes it possible to analytically derive the stationary distribution \cite{Van1992}.
Then following the additivity, we obtain the results in the more complicated heterogeneous cases, which greatly reduces the computational complexity.
As pointed out by recent studies \cite{Amaral2015, Amaral2016}, a heterogeneous system may be also well approximated by the corresponding homogeneous system with averaged parameters.
Although it looks similar to our results, it is different from the additivity we revealed since i) the additivity property connects a heterogeneous system with its different homogeneous counterparts rather than one homogeneous system with averaged parameters;
ii) the parameters such as payoff entries \cite{Amaral2015, Amaral2016} are suitable for averaging whereas update functions would seem inappropriate for this operation.

In addition, in the limit of weak selection, we analytically obtain a condition to tell whether one strategy is more abundant than the other in the steady state.
This condition coincides with that derived under a homogeneous aspiration-based update rule in finite populations \cite{Du2014}.
The meaning of this finding is twofold: it reveals that the heterogeneity of aspiration-based rules does not affect the condition; it shows the consistency between finite and infinite populations for evolutionary dynamics induced by aspiration-based rules while there seems to be an inconsistency for the dynamics with imitation-based ones \cite{Traulsen2005, Gokhale2010}.

Beyond weak selection, we show that our equations accurately predict the final evolutionary outcomes in public goods games, especially in well-mixed populations.
In structured populations, deviations occur when the selection intensity becomes very strong.
Despite of this, the predictions match reasonably well with simulations.
This is unexpected since (i) pair approximations neglect the triplet correlations, which affects the payoffs of individuals in multi-player games;
(ii) small deviations in payoffs may dramatically change individuals' strategic behavior under strong selection intensities.
The good agreement with simulations suggests that our work offers an efficient method to calculate the evolutionary outcomes for strong selections under heterogeneous aspiration dynamics.

Besides, due to the generality of our formalism, the framework we present can be readily applied to other social dilemmas and other combinations of update functions.
It may also handle the situations where the population consists of individuals using different types of update rules, for instance, imitation-based and aspiration-based ones.
Thus, our work provides a general approach to address the effect of heterogeneity in update rules on evolutionary outcomes.

\begin{acknowledgments}
    L.Z. and L.W. are supported by NSFC (Grants No. 61751301 and No. 61533001).
    B.W. is grateful for funding by the NSFC (Grant No. 61603049 and No. 61751301) and the Fundamental Research Funding for the Central Universities in China (No. 2017RC19).
    L.Z. acknowledges the support from China Scholarship Council (No. 201606010270) and the Levin Lab.
    V.V.V. acknowledges the support by US Defense Advanced Research Projects Agency (D17AC00005), by the National Science Foundation grant GEO-1211972, and by Funda\c{c}\~{a}o para a Ci\^{e}ncia e a Tecnologia (FCT), Portugal through grants PTDC/MAT-STA/3358/2014 and PTDC/EEI-SII/5081/2014.
\end{acknowledgments}
\appendix
\section{Well-mixed population}\label{AppendixWellMixed}
In well-mixed populations, we denote the number of $A$-players using update function $i$ as $n_i$ ($0 \leq n_i \leq N_i$, $i=1,2$).
At each time step, an individual is randomly selected from the population and it switches its strategy with a probability given by its own update function.
The resulting evolutionary dynamics can be described \textit{exactly} by a two-dimensional birth-death process with reflecting states \cite{Van1992}.
Under this process, if a $B$-player using update function $g_1$ is selected and it changes to an $A$-player, the state variable $n_1$ will increase by 1.
The transition probability associated with this event is
\begin{equation*}
 T_1^+(n_1, n_2) = \frac{N_1-n_1}{N}g_1\left(\beta(e-\pi_B(n_1,n_2))\right),
\end{equation*}
where $\pi_B(n_1,n_2) = {{N-1}\choose {d-1}}^{-1}\sum_{k=0}^{d-1}{{n_1+n_2}\choose {k}} {{N-n_1-n_2-1}\choose {d-1-k}}b_k$.
Similarly, all the other transition probabilities are
\begin{eqnarray*}
  T_1^-(n_1, n_2) &=& \frac{n_1}{N}g_1\left(\beta(e-\pi_A(n_1,n_2))\right),  \\
  T_2^+(n_1, n_2) &=& \frac{N_2-n_2}{N}g_2\left(\beta(e-\pi_B(n_1,n_2))\right),  \\
  T_2^-(n_1, n_2) &=& \frac{n_2}{N}g_2\left(\beta(e-\pi_A(n_1,n_2))\right),
\end{eqnarray*}
and $T_{1,2}^0(n_1, n_2) = 1 - T_1^+(n_1, n_2) - T_1^-(n_1, n_2) - T_2^+(n_1, n_2) - T_2^-(n_1, n_2)$, which correspond to the event of $n_1$ decreases by 1, $n_2$ increases by 1, $n_2$ decreases by 1, and both $n_1$ and $n_2$ do not change, respectively.
Here, $\pi_A(n_1,n_2) = {{N-1}\choose {d-1}}^{-1}\sum_{k=0}^{d-1}{{n_1+n_2-1}\choose {k}} {{N-n_1-n_2}\choose {d-1-k}}a_k$.
Determined by these transition probabilities, this process admits a unique stationary distribution where the system can be in every possible state with a positive probability \cite{Van1992, Grinstead2012}.

Denoting the probability in state $(n_1, n_2)$ at time $\tau$ as $P^{\tau}(n_1, n_2)$, we can write the master equation as
\begin{eqnarray*}
P^{\tau+1}(n_1, n_2) - P^{\tau}(n_1, n_2) &=& P^{\tau}(n_1-1, n_2)T_1^+(n_1-1, n_2) + P^{\tau}(n_1+1, n_2)T_1^-(n_1+1, n_2) \\
  & &+P^{\tau}(n_1, n_2-1)T_2^+(n_1, n_2-1) + P^{\tau}(n_1, n_2+1)T_2^-(n_1, n_2+1) \\
  & &-  P^{\tau}(n_1, n_2)\left(T_1^+(n_1, n_2) + T_1^-(n_1, n_2)+T_2^+(n_1, n_2)+T_2^-(n_1, n_2)\right).
\end{eqnarray*}
Now we introduce the notations $y_1=n_1/N \in [0,z_1]$, $y_2=n_2/N\in[0,z_2]$, $t=\tau/N$ and the probability density $\rho(y_1,y_2,t) = N^2 P_{\tau}(n_1,n_2)$.
When $N \gg 1$, following the similar procedure to that in \cite{Traulsen2006, Pacheco2014, Vasconcelos2017}, we obtain the Langevin equations
\begin{eqnarray*}
  \dot{y}_1 &=& a_1(y_1,y_2) + b_1(y_1,y_2)\xi_1, \\
  \dot{y}_2 &=& a_2(y_1,y_2) + b_2(y_1,y_2)\xi_2,
\end{eqnarray*}
where $a_i(y_1, y_2) = T_i^+(y_1,y_2) - T_i^-(y_1,y_2)$, $b_i(y_1, y_2) = \sqrt{\left[T_i^+(y_1,y_2) + T_i^-(y_1,y_2)\right]/N}$, and $\xi_i$ is uncorrelated Gaussian white noise with unit variance (here, $i=1, 2$).
Note that in the above equations, the evolution of $y_1$ ($y_2$) is only affected by noise $\xi_1$ ($\xi_2$), which is different from those obtained for a homogeneous population with multiple strategies \cite{Traulsen2006, Vasconcelos2017}.

As $N\rightarrow \infty$, the diffusion term $b_i$ will vanish with $1/\sqrt{N}$ and we approximate the hypergeometric distribution by the binomial distribution.
Then we get the set of deterministic Eqs.~(\ref{WellMixedODE}) in the main text.

Under the limit of weak selection $\beta \rightarrow 0$, based on Eqs.~(\ref{NormalizedWellMixedODE}), we obtain
\begin{equation}\label{ODE_infinity}
    \dot{x}_i = (1-2x_i)g_i(0)+(1-2x_i)g'_i(0)e\beta + \left[x_i\pi_A-(1-x_i)\pi_B \right]g'_i(0)\beta + O(\beta^2).
\end{equation}
When the system reaches its steady state, $x_i$ is approximately $1/2$ plus some deviation which is of the first order of selection intensity. Here, we denote the fraction of $A$-players among all the players using update function $g_i$ at the steady state as
\begin{equation}\label{steadyStateApprox}
  x_i^* = \frac{1}{2}+\delta_i\beta + O(\beta^2),
\end{equation}
thus
\begin{equation}
  \pi_A = \frac{1}{2^{d-1}}\sum_{k=0}^{d-1}{{d-1}\choose k}a_k + O(\beta),
  ~~\pi_B = \frac{1}{2^{d-1}}\sum_{k=0}^{d-1}{{d-1}\choose k}b_k + O(\beta). \label{ApproxPayoffs}
\end{equation}
Let the left-hand side of Eqs.~(\ref{ODE_infinity}) equal to zero.
Inserting Eqs.~(\ref{ApproxPayoffs}) and (\ref{steadyStateApprox}) into the resulting equations, we get
\begin{equation*}
  0 = -2\delta_i\beta g_i(0) + \frac{g'_i(0)\beta}{2^d}\sum_{k=0}^{d-1}{{d-1}\choose k}(a_k-b_k)+O(\beta^2),
\end{equation*}
which leads to
\begin{equation*}\label{firstOrderDeviation}
  \delta_i = \frac{1}{2^{d+1}}\frac{g'_i(0)}{g_i(0)}\sum_{k=0}^{d-1}{{d-1}\choose k}(a_k-b_k).
\end{equation*}
Then the average abundance of strategy $A$ in the whole population in the steady state is
\begin{equation*}
  x^* = z_1x_1^* + z_2x_2^* = \frac{1}{2}+ \frac{1}{2^{d+1}}\left(z_1\frac{g'_1(0)}{g_1(0)}+z_2\frac{g'_2(0)}{g_2(0)}\right)\sum_{k=0}^{d-1}{{d-1}\choose k}(a_k-b_k)\beta + O(\beta^2).
\end{equation*}

\section{Structured population}\label{AppendixStructured}
Based on the pair approximation \cite{Ohtsuki2006}, we tailor it for our heterogenous aspiration dynamics and obtain
\begin{eqnarray*}
  & &p_{A_i}+p_{B_i} = 1 \\
  & &p_A + p_B = 1 \\
  & &p_A = z_1p_{A_1}+z_2p_{A_2} \\
  & &p_{YZ} = p_{ZY} \\
  & &p_{YZ} = q_{Y|Z}\cdot p_Z = q_{Z|Y}\cdot p_Y \\
  & &q_{A|Y} + q_{B|Y} = 1
\end{eqnarray*}
where $i=1,2$ and $Y, Z=A,B$ (see the explanation of these notations in the main text).
Note that $p_{YZ}$ is the pair pointing from $Y$ to $Z$.
In this sense, $p_{YY}$ will be counted twice.
From these equations, we can get that all the quantities listed above are functions of $p_{A_1}$, $p_{A_2}$, and $q_{A|A}$.
This implies that the whole system can be described by $p_{A_1}$, $p_{A_2}$, and $q_{A|A}$.

In structured populations, the average payoff of a focal $A$-player with $k_A$ neighbors playing strategy $A$ is $\pi_{A, k_A} = (k+1)^{-1}[a_{k_A}+k_{A}\sum_{l_A=0}^{k-1}{k-1 \choose l_A}q_{A|A}^{l_A}q_{B|A}^{k-1-l_A}a_{l_A+1}+k_{B}\sum_{l_A=0}^{k-1}{k-1 \choose l_A}q_{A|B}^{l_A}q_{B|B}^{k-1-l_A}a_{l_A}]$.
Similarly, for a focal $B$-player with the same neighbor configuration, its average payoff is $\pi_{B, k_A} = (k+1)^{-1}[b_{k_A}+k_{A}\sum_{l_A=0}^{k-1}{k-1 \choose l_A}q_{A|A}^{l_A}q_{B|A}^{k-1-l_A}b_{l_A+1} +k_{B}\sum_{l_A=0}^{k-1}{k-1 \choose l_A}q_{A|B}^{l_A}q_{B|B}^{k-1-l_A}b_{l_A}]$.

Based on the above equations, the probability for $A_1$ individuals to increase (decrease) by $1/(z_1N)$ is
\begin{equation}
  \begin{split}
  & {\rm Pr}\left(\Delta p_{A_1} = \frac{1}{z_1N}\right) = z_1p_{B_1}\sum_{k_A=0}^k {k\choose k_A} q_{A|B}^{k_A} q_{B|B}^{k_B}g_1\left(\beta(e-\pi_{B, k_A})\right), \\
  & {\rm Pr}\left(\Delta p_{A_1} = -\frac{1}{z_1N}\right) = z_1p_{A_1}\sum_{k_A=0}^k {k \choose k_A} q_{A|A}^{k_A} q_{B|A}^{k_B}g_1\left(\beta(e-\pi_{A, k_A})\right).
  \end{split}
  \label{ProbPA1}
\end{equation}
Also, we can write the probability for $A_2$ individuals to increase (decrease) by $1/(z_2N)$ as
\begin{equation}
\begin{split}
 & {\rm Pr}\left(\Delta p_{A_2} =\frac{1}{z_2N}\right) = z_2p_{B_2}\sum_{k_A=0}^k {k \choose k_A} q_{A|B}^{k_A} q_{B|B}^{k_B}g_2\left(\beta(e-\pi_{B, k_A})\right), \\
 & {\rm Pr}\left(\Delta p_{A_2} =-\frac{1}{z_2N}\right) = z_2p_{A_2}\sum_{k_A=0}^k {k \choose k_A} q_{A|A}^{k_A} q_{B|A}^{k_B}g_2\left(\beta(e-\pi_{A, k_A})\right).
\end{split}
  \label{ProbPA2}
\end{equation}
Then the expected rate of change for $p_{A_1}$ ($p_{A_2}$) is written as
\begin{eqnarray}
  \dot{p}_{A_1} &=& \frac{1}{z_1N}{\rm Pr}\left(\Delta p_{A_1}=\frac{1}{z_1N}\right)-\frac{1}{z_1N}{\rm Pr}\left(\Delta p_{A_1}=-\frac{1}{z_1N}\right), \label{Appendix:DpA1}\\
  \dot{p}_{A_2} &=& \frac{1}{z_2N}{\rm Pr}\left(\Delta p_{A_2}=\frac{1}{z_2N}\right)-\frac{1}{z_2N}{\rm Pr}\left(\Delta p_{A_2}=-\frac{1}{z_2N}\right). \label{Appendix:DpA2}
\end{eqnarray}
Similarly, for the dynamics of the pairs, we could write
\begin{equation}\label{Appendix:DpAA}
  \dot{p}_{AA} = ~\sum_{k_A=0}^{k}
  \left(\frac{2k_A}{kN}\right){\rm Pr}\left(\Delta p_{AA}=\frac{2k_A}{kN}\right) +\sum_{k_A=0}^k
  \left(-\frac{2k_A}{kN}\right){\rm Pr}\left(\Delta p_{AA}=-\frac{2k_A}{kN}\right),
\end{equation}
where
\begin{equation}\label{ProbPAA}
\begin{split}
& {\rm Pr}\left(\Delta p_{AA}=\frac{2k_A}{kN}\right) = {k \choose k_A} q_{A|B}^{k_A} q_{B|B}^{k_B}\left[z_1p_{B_1}g_1\left(\beta(e-\pi_{B, k_A})\right)+z_2p_{B_2}g_2\left(\beta(e-\pi_{B, k_A})\right) \right], \\
& {\rm Pr}\left(\Delta p_{AA}=-\frac{2k_A}{kN}\right) = {k \choose k_A} q_{A|A}^{k_A} q_{B|A}^{k_B}\left[z_1p_{A_1}g_1\left(\beta(e-\pi_{A, k_A})\right)+z_2p_{A_2}g_2\left(\beta(e-\pi_{A, k_A})\right) \right].
\end{split}
\end{equation}

Based on the relation $q_{A|A}=\frac{p_{AA}}{p_A}$ and $p_A=z_1p_{A_1}+z_2p_{A_2}$, we have
\begin{equation}\label{Appendix:DqAA}
  \dot{q}_{A|A}=\frac{d}{dt}\left(\frac{p_{AA}}{p_A}\right) = \frac{1}{p_A}\left(\dot{p}_{AA}-q_{A|A}(z_1\dot{p}_{A_1}+z_2\dot{p}_{A_2})\right).
\end{equation}

Substituting Eqs.~(\ref{ProbPA1}), (\ref{ProbPA2}) and (\ref{ProbPAA}) into (\ref{Appendix:DpA1}-\ref{Appendix:DqAA}), we obtain the closed dynamical system described by Eqs.~(\ref{DpA1NewMain}-\ref{DqAANewMain}).

In order to analyze the dynamics under the weak selection limit $\beta \rightarrow 0$,
we first set $\beta = 0$ and let the left-hand side of the above equations equal to zero. Then for this unperturbed system, we obtain $p_{A_i}^*=p_{B_i}^*=\frac{1}{2}$ ($i=1,2$) and $q_{A|A}^*=\frac{1}{2}$ in the steady state.
As the selection intensity $\beta \rightarrow 0$, using perturbation theory \cite{Khalil2002} we have the approximations $p_{A_i}^*=\frac{1}{2}+\epsilon_i\beta +O(\beta^2)$ ($i=1,2$) and $q_{A|A}^*=\frac{1}{2}+O(\beta)$.
Moreover, as $\beta \rightarrow 0$, when the system reaches its steady state, Eqs.~(\ref{DpA1NewMain}-\ref{DpA2NewMain}) become
\begin{align*}
&0 = \frac{1}{N}g_1(0)(p_{B_1}^*-p_{A_1}^*) \\
&+\frac{g'_1(0)}{N}\beta\left[e(p_{B_1}^*-p_{A_1}^*)
  + p_{A_1}^*\sum_{k_A=0}^k{k \choose k_A}(q_{A|A}^*)^{k_A}(q_{B|A}^*)^{k_B}\pi_{A,k_A}^*
  -p_{B_1}^*\sum_{k_A=0}^k{k \choose k_A}(q_{A|B}^*)^{k_A}(q_{B|B}^*)^{k_B}\pi_{B,k_A}^*\right] \\
&+ O(\beta^2), \\
&0 = \frac{1}{N}g_2(0)(p_{B_2}^*-p_{A_2}^*) \\
& +\frac{g'_2(0)}{N}\beta\left[e(p_{B_2}^*-p_{A_2}^*)
  + p_{A_2}^*\sum_{k_A=0}^k{k \choose k_A}(q_{A|A}^*)^{k_A}(q_{B|A}^*)^{k_B}\pi_{A,k_A}^*
  -p_{B_2}^*\sum_{k_A=0}^k{k \choose k_A}(q_{A|B}^*)^{k_A}(q_{B|B}^*)^{k_B}\pi_{B,k_A}^*\right]\\
&  + O(\beta^2).
\end{align*}
Inserting the approximations for $p_{A_i}^*$ and $q_{A|A}^*$ into the above equations and neglecting the higher-order terms of $\beta$, we obtain $\epsilon_i = \frac{1}{2^{d+1}}\frac{g'_i(0)}{g_i(0)}\sum_{k=0}^{d-1}{{d-1}\choose k}(a_k-b_k)$.
This leads to the same average abundance of strategy $A$ in the whole population (here, $x^* = z_1p_{A_1}^*+z_2p_{A_2}^*$) as that in the well-mixed population (see Appendix \ref{AppendixWellMixed}).

\section{Higher-order approximations for the additivity}\label{HigherOrderApprox}
In \ref{AppendixWellMixed}, we derive the first-order approximation of $x_i^*$ and $x^*$.
It validates the additivity of heterogeneous aspiration dynamics under the limit of weak selection.
Here, to offer an intuitive explanation on why the additivity applies to strong selections for public goods games in well-mixed populations, we implement higher-order approximations.
Setting the left-hand side of equations (\ref{NormalizedWellMixedODE}) equal to zero and rearranging the items, we have that all the fixed points satisfy the following implicit equations
\begin{equation*}
    x_i^* = \frac{1}{1+\frac{g_i(\beta(e-\pi_A(x_1^*,x_2^*)))}{g_i(\beta(e-\pi_B(x_1^*,x_2^*)))}}, ~~~i=1,2.
\end{equation*}
Based on perturbation theory \cite{Khalil2002}, we construct finite Taylor series in $\beta$ at $\beta=0$ (up to the third-order) for both the left-hand and right-rand side of the above equations.
By matching the coefficients of the same power of $\beta$, we can calculate the higher-order approximation for $x_i^*$.
The calculation proceeds as follows:
\begin{enumerate}
  \item For the left-hand side, $x_i^*=x_{i0}+x_{i1}\beta+x_{i2}\beta^2 + x_{i3}\beta^3 + O(\beta^4)$;
  \item For the right-hand side, $F_i(x_1,x_2)=\left[1+\frac{g_i(\beta(e-\pi_A(x_1,x_2)))}{g_i(\beta(e-\pi_B(x_1,x_2)))}\right]^{-1}=\sum_{l=0}^2\left.
      \frac{\partial^l F_i(x_1,x_2)}{\partial \beta^l}\right|_{\beta=0}\beta^l + O(\beta^3)$;
  \item Insert the approximation of $x_i^*$ into that of $F_i(x_1,x_2)$ and neglect the higher-order terms of $\beta^3$;
  \item Match the constant term and coefficients of the same power of $\beta$ from both sides and calculate $x_{i0}, x_{i1}, x_{i2}$, and $x_{i3}$.
\end{enumerate}
Following similar procedures, we can obtain the third-order approximations for $x^*_{\text{I}}$ and $x^*_{\text{II}}$.
After that, we calculate the deviation of additivity $\mathcal{D}_{\text{MH}}=x^*_{\text{mixed}}-z_1x^*_{\text{I}}-z_2x^*_{\text{II}}=z_1x^*_1+z_2x^*_2-z_1x^*_{\text{I}}-z_2x^*_{\text{II}}= \mathcal{D}^{(0)}_{\text{MH}}+\mathcal{D}^{(1)}_{\text{MH}}\beta
+\mathcal{D}^{(2)}_{\text{MH}}\beta^2 + \mathcal{D}^{(3)}_{\text{MH}}\beta^3 + O(\beta^4)$, where $\mathcal{D}^{(l)}_{\text{MH}}$ is the coefficient associated with $\beta^l$.

As already shown in the main text, $\mathcal{D}^{(0)}_{\text{MH}}=0$, and  $\mathcal{D}^{(1)}_{\text{MH}}=0$. For higher-order approximations, we obtain
\begin{eqnarray*}
  \mathcal{D}^{(2)}_{\text{MH}} &=& -\frac{1}{16}z_1z_2(d-1)R_1R_2G_1^2, \\
  \mathcal{D}^{(3)}_{\text{MH}} &=& \frac{1}{64}(d-1)^2 R_1 R_2^2 \left(G_3^3-G_5\right) -\frac{1}{16}z_1 z_2 (d-1) G_1 G_2 R_1 R_2 (2e - S_1)\\
  & &- \frac{1}{128}z_1z_2(d-1)(d-2) G_1^2 (G_3+G_4) R_1^2 R_3 +\frac{1}{32}z_1z_2(d-1)G_1 G_2 R_1^2 S_2,
\end{eqnarray*}
where
\begin{eqnarray*}
R_1 &=& \frac{1}{2^{d-1}}\sum_{k=0}^{d-1}{d-1 \choose k}(a_k-b_k), ~~S_1 = \frac{1}{2^{d-1}}\sum_{k=0}^{d-1}{d-1 \choose k}(a_k+b_k),  \\
R_2 &=& \frac{1}{2^{d-2}}\sum_{k=0}^{d-2}{d-2 \choose k}(a_{k+1}-b_{k+1}-a_k+b_k),
  ~~S_2 = \frac{1}{2^{d-2}}\sum_{k=0}^{d-2}{d-2 \choose k}(a_{k+1}+b_{k+1}-a_k-b_k), \\
R_3 &=& \frac{1}{2^{d-3}}\sum_{k=0}^{d-3}{d-3 \choose k}(a_{k+2}-b_{k+2}-2a_{k+1}+2b_{k+1}+a_k-b_k), \\
G_1 &=& \frac{g_1'(0)}{g_1(0)}-\frac{g_2'(0)}{g_2(0)}, ~~G_2 = \left(\frac{g_2'(0)}{g_2(0)}\right)^2-\left(\frac{g_1'(0)}{g_1(0)}\right)^2
  + \frac{g_1''(0)}{g_1(0)} - \frac{g_2''(0)}{g_2(0)},  \\
G_3 &=& z_1\frac{g_1'(0)}{g_1(0)}+z_2\frac{g_2'(0)}{g_2(0)},~~G_4 = \frac{g_1'(0)}{g_1(0)}+\frac{g_2'(0)}{g_2(0)}, ~~G_5 = z_1\left(\frac{g_1'(0)}{g_1(0)}\right)^3+z_2\left(\frac{g_2'(0)}{g_2(0)}\right)^3.
\end{eqnarray*}
In particular, for public goods games with payoff entries $a_k=rc(k+1)/d-c$ and $b_k=rck/d$ ($k=0,1,\cdots,d-1$), $R_1=rc/d-c$, $R_2=0$, $R_3=0$, $S_2=2rc/d$.
The deviation of additivity $\mathcal{D}_{\text{MH}}$ is simplified to
\begin{equation}\label{DeviationHigherOrderApprox}
  \mathcal{D}_{\text{MH}} = \frac{1}{16}z_1z_2(d-1)\frac{rc}{d}\left(\frac{rc}{d}-c\right)^2G_1G_2\beta^3 + O(\beta^4).
\end{equation}
This means that the deviation from perfect additivity only occurs at the third-order approximation of $\beta$, which explains why the additivity in public goods games is robust to selection intensities.


%

\end{document}